\documentclass[english,aps,twocolumn]{revtex4}
\usepackage{epsfig}
\usepackage{graphicx}
\usepackage{amsmath}
\makeatletter
\@ifundefined{textcolor}{}
{
 \definecolor{BLACK}{gray}{0}
 \definecolor{WHITE}{gray}{1}
 \definecolor{RED}{rgb}{1,0,0}
 \definecolor{GREEN}{rgb}{0,1,0}
 \definecolor{BLUE}{rgb}{0,0,1}
 \definecolor{CYAN}{cmyk}{1,0,0,0}
 \definecolor{MAGENTA}{cmyk}{0,1,0,0}
 \definecolor{YELLOW}{cmyk}{0,0,1,0}
 }

\makeatother

\usepackage{babel}

\begin{document}

\title{On the existence of Rydberg nuclear molecules}

\author{C. A. Bertulani}

\affiliation{Department of Physics and Astronomy, Texas A\&M University-Commerce, Commerce, TX 75429-3011, USA}
\affiliation{Institut f\"ur Kernphysik, Technische Universit\"at Darmstadt, D-64289 Darmstadt, Germany}

\author{T. Frederico}

\affiliation{Departamento de F\'{i}sica, Instituto Tecnol\'{o}gico de Aeron\'{a}utica, 12228-900 S\~{a}o Jos\'{e} dos Campos, SP, Brazil}

\author{M. S. Hussein}

\affiliation{Instituto Tecnol\'{o}gico de Aeron\'{a}utica, DCTA,12.228-900 S\~{a}o Jos\'{e} dos Campos, SP, Brazil}
\affiliation{Instituto de Estudos Avan\c{c}ados, Universidade de S\~{a}o Paulo C. P.
72012, 05508-970 S\~{a}o Paulo-SP, Brazil}
\affiliation{Instituto de F\'{\i}sica, Universidade de S\~{a}o Paulo, C. P. 66318, 05314-970 S\~{a}o Paulo, SP, Brazil}

\begin{abstract}
Present nuclear detection techniques prevents us from determining if the analogue of a Rydberg molecule exists for the nuclear case. But nothing in nature disallows their existence. As in the atomic case, Rydberg nuclear molecules would be a laboratory for new aspects and applications of nuclear physics. We propose that Rydberg nuclear molecules, which represent the exotic, halo nuclei version, such as $^{11}$Be + $^{11}$Be, of the well known quasimolecules observed in stable nuclei such as $^{12}$C + $^{12}$C, might be common structures that could manifest their existence along the dripline. A study of possible candidates and the expected structure of such exotic clustering of two halo nuclei: the Rydberg nuclear molecules, is made on the basis of three different methods. It is shown that such cluster structures might be stable and unexpectedly common.
\end{abstract}

\maketitle

In atomic physics, a Rydberg atom is an excited state of the atom with one or more electrons having a very high principal quantum number, $n$ ($\approx$ 40), forming a weakly-bound state \cite{Ga94}. They are characterized by a large spatial extent, usually a few thousand times larger than an atom in its ground state.  These atoms exhibit interesting phenomena such as a large response to electric and magnetic fields, and classical orbital behavior.  The inner electrons shield the outer electron so that the electric potential of the outer electron(s) look identical to that of a hydrogen atom.  Among other applications in physics, Rydberg atoms  are common in interstellar space, and  are an important radiation source for astronomers. They are also common in plasmas because of the recombination of electrons and positive ions yielding stable Rydberg atoms. Rydberg atoms are also strong candidates to realize a quantum computer \cite{Ja00}. 

Recently, the creation of atomic Rydberg molecules, or Rydberg dimers,  was proposed \cite{Gr00}, and verified experimentally by \cite{ Ov09, Ve09}. These dimers were understood as the result of the binding that the Rydberg electron in a Rydberg atom binds with another atom in its ground state or with another Rydberg atom.
These, critically stable, loosely bound diatomic systems are quite extended with a molecular radius, $r_{RM}$ that could reach several thousands Bohr radii  ($r_{RM} \approx 2000 a_{0} = 1 \times 10^{3}{\buildrel _{\circ} \over {\mathrm{A}}} = 1 \times 10^{8}$ fm). The binding energy in the case of $^{87}$Rb(35s)+$^{87}$Rb studied in \cite{Gr00, Ov09, Ve09}, is about $E_{RM} \approx 1 \times 10^{-7}$ eV. Much more recently, more exotic butterfly Rydberg  molecules, have been observed \cite{BRM1}.

It is natural to ask whether such molecular states could exist in nuclei. The type of nuclei that could resemble Rydberg atoms are the halo ones, where ,e.g., a neutron or two neutrons occupy a halo orbit far a way from the center of the core. Can two such halo nuclei, when they are allowed to interact, form a Rydberg nuclear molecule? In the following we supply a tentative answer in the affirmative, but these critically stable nuclear dimers are much less dramatic than their atomic counter parts.

Nuclear physics has been revolutionized by the physics of loosely-bound, dripline nuclei \cite{BCH93}. Much of the present experimental and theoretical efforts in nuclear physics has been directed to this area, with applications to several areas of science, ranging from astrophysics, cosmology, nuclear medicine and national security. Exotic nuclei are ideal laboratories for mesoscopic physics, having lead science to the development of new and poorly explored phenomena such as the Efimov effect \cite{Ef70} (see also \cite{Fre12}). But a Rydberg nucleus is an elusive object in nuclear physics.  Highly excited nuclei posses numerous decay channels,  allowing for very fast decays either by  particle, $\gamma$, or $\beta$-emission. Decay times are very short for the most loosely-bound nuclei, which poses a practical  hurdle for their detection. Nearly stable, ground-states of known loosely bound nuclei, e.g., $^{11}$Li, also have a matter distribution which is only up to a factor of 2 (based on root-mean-square radii) larger than neighboring isotopes. This contrasts to a factor of 1000 for the ratio between sizes of Rydberg atoms to typical atoms \cite{Ga94}.

In this letter we propose that Rydberg nuclear molecules might be seen in nuclear molecular structures and clustering in composite light halo or Borromean nuclei, which are more stable and amenable to measurements. This argument is based on the simple fact that bound unstable Borromean systems benefit from three- and four-body correlations, such as in$^{11}$Li, a nucleus for which none of the subsystems $^{10}$Li+n, or n+n, exist, but which is bound in the long range effective potential $\propto -\rho^{-2}$, where $\rho$ is the hyper-radius of the three-particle system, as shown by Efimov in the case of three identical bosons \cite{Ef70}.  Weakly bound four-body states close to an Efimov trimer state \cite{Pla04,Yam06,St09,Had11} were evidenced theoretically and  experimentally for the first time in a cold atom setup by the Innsbruck group \cite{Fer09}. That makes plausible  that such states might be achievable also  with nuclear molecules.

Clusterization is a well-known phenomenon in  nuclear physics. Usual nuclear models for light nuclei assume the existence of a core nucleus surrounded by nucleons, such as  $^{11}$Li = $^9$Li+n+n, or $^8$He = $^4$He+n+n+n+n. The internal structure of the core nucleus is accounted for by energy, angular momentum, and Pauli projection of the nuclear state of interest. Assuming that they exist, Rydberg nuclear molecules would be sensitive to the tail of the nuclear interaction, where the bulk of the matter distribution of valence nucleon extends way beyond the core. For such a system, one can build a nuclear effective potential using the idea of a pseudo-potential, which was first introduced by E. Fermi to describe the scattering of a free neutron by a nucleus \cite{Fe36}. 

The pseudo-potential is an attempt to replace the complicated effects of the motion of the core (i.e. non-valence) particles  by an effective potential such that core states are eliminated. Originally, the pseudo-potential was used to describe the scattering of a low-energy electron from a neutral atom \cite{Fe36}. In atomic physics, the polarization potential for the electron-atom interaction is always attractive, but the s-wave scattering can give rise to either a positive or a negative scattering length depending on the relative phase between the ingoing and the scattered electron waves. This was explored in Ref. \cite{Gr00} to predict novel molecular binding  from a low-energy Rydberg electron scattering from an atom with negative scattering length. The scattering is assumed to be s-wave scattering, and therefore spherically symmetric. Hence, the potential is given as a function of radius, $V(R)=2\pi a \hbar^2\delta(R)/m$, where $a$ is the neutron-nucleus scattering length and $m$ the nucleon mass. Averaged over many scattering events and weighted with the local  density $|\psi(R)|^2$,  Fermi's method  leads to an effective potential between the scattering partners given by
\begin{equation}
V(R)= {2\pi \hbar^2 \over m} a(k) \left|\psi(R)\right|^2. \label{pseudo}
\end{equation}

\begin{figure}[ht]
{\par\centering \resizebox*{0.45\textwidth}{!} {\includegraphics{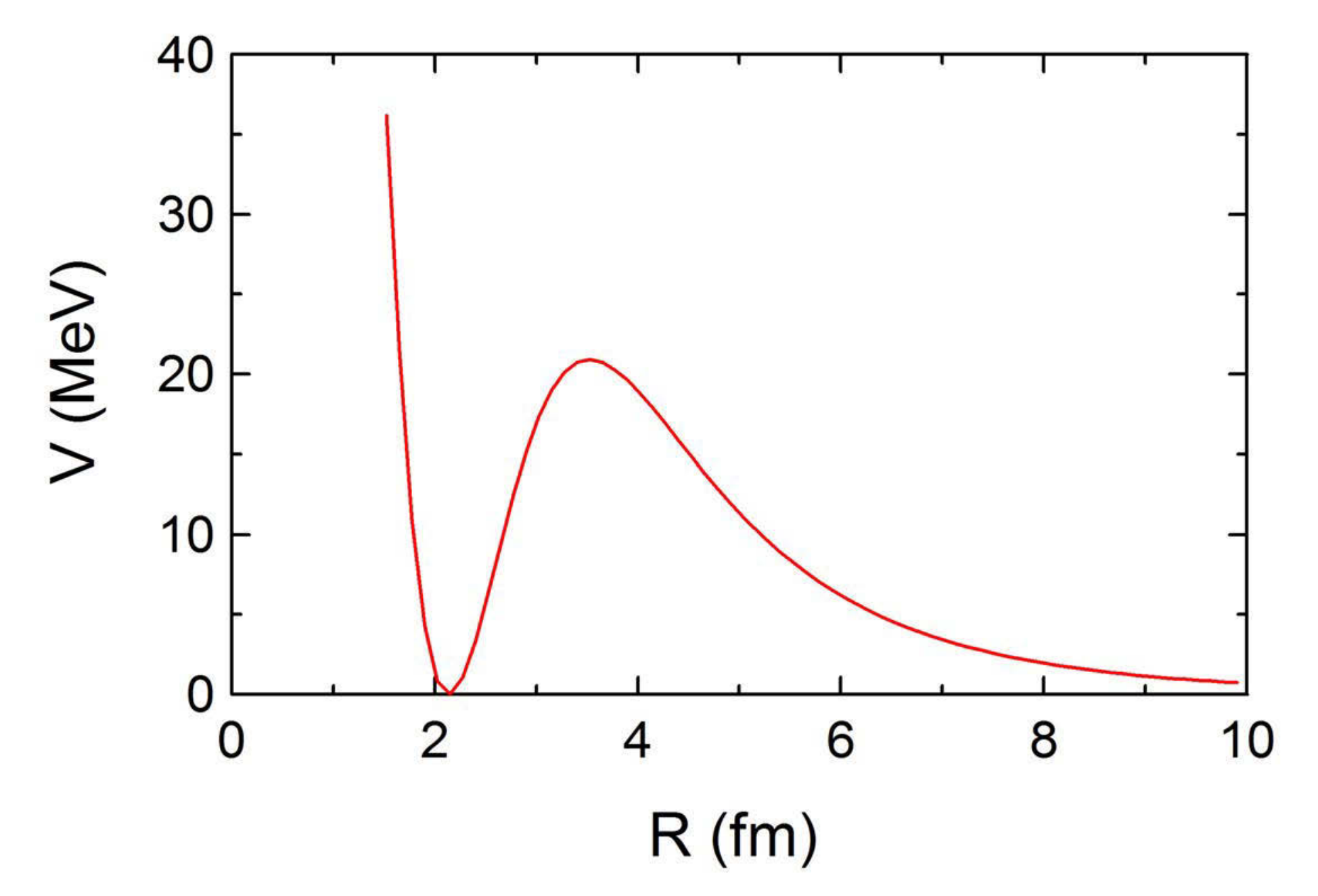}}\par}
\caption{\label{fig1} (Color online) Pseudo-potential for the n+$^{10}$Be system.}
\end{figure}

Tentatively, we can apply this result for neutron+$^{10}$Be. Here, the ground state is a $2s_{1/2}$ due to parity inversion and we use the  wave function $\psi_{2s_{1/2}}(R)$ of a valence neutron coupled to a $0^+$ $^{10}$Be core from a simple Woods-Saxon plus spin-orbit potential model, with potential depth adjusted to reproduce the binding energy of 504 keV.  The strength of the spin-orbit term was adjusted by requiring that the first excited (${1\over 2}^-$) state in $^{11}$Be be at the experimental  excitation energy of 320 keV. For the $a(k)$ we use the value $a_0\equiv a(0)=6.88$ fm for the neutron-$^{10}$Be scattering length. We obtain the result in Figure \ref{fig1}. The potential shows a pocket within a close range from the core and a long tail because of the low binding energy. The potential pocket displayed in Figure \ref{fig1} suggests that a resonance in the scattering of a neutron by the $^{11}$Be system might emerge through this mechanism. Had we used an artificially constructed loosely bound state in $^{11}$Be for $\psi({\bf r)}$, the pocket would still be visible at a larger distance from the n + $^{10}$Be system because in this oversimplified model the pocket just reflects the node in the wave function. As with the proposed Rydberg molecules \cite{Gr00}, the additional n + $^{10}$Be system with a long tail for the neutron wavefunction, could become momentarily or permanently localized in this pocket.  In particular,  large $n$ nuclear-equivalent of Rydberg states in n + A systems would enhance this possibility.   Such Rydberg states would consist of very loosely bound states in halo nuclei.

The application of the pseudo-potential to a nuclear molecular state in, e.g., $^{10}$Be+n+$^{10}$Be, or $^{10}$Be+n+n+$^{10}$Be, might not work, as the principal quantum number, $n=2$, is too small to justify the validity of Eq. \eqref{pseudo}.  But  Eq. \eqref{pseudo} is applicable if a molecular state is reminiscent of a large $n$, resonant state, in $^{10}$Be+n. Such hypothesis is hard to check without experimental measurements of excited states in  nuclei such as $^{20}$C ($^{9}$Li+n+n+$^9$Li), or $^{22}$O ($^{10}$Be+n+n+$^{10}$Be), perhaps the best candidates for  hyper-molecular states. A large principal quantum number $n$ of the 4-body molecular system can be assumed for a molecular state.  Such hypothesis, was used by Greene et al. \cite{Gr00} to predict the existence of Rydberg atomic molecules which attracted a great interest in atomic physics. In nuclear systems the existence of such molecules cannot be ruled out either. While the last electron orbital in Rydberg atoms experiences a shielding of the nuclear charge by the other electrons, no such mechanism exists with nuclear systems.  
   
\begin{figure}[ht]
{\par\centering \resizebox*{0.45\textwidth}{!} {\includegraphics{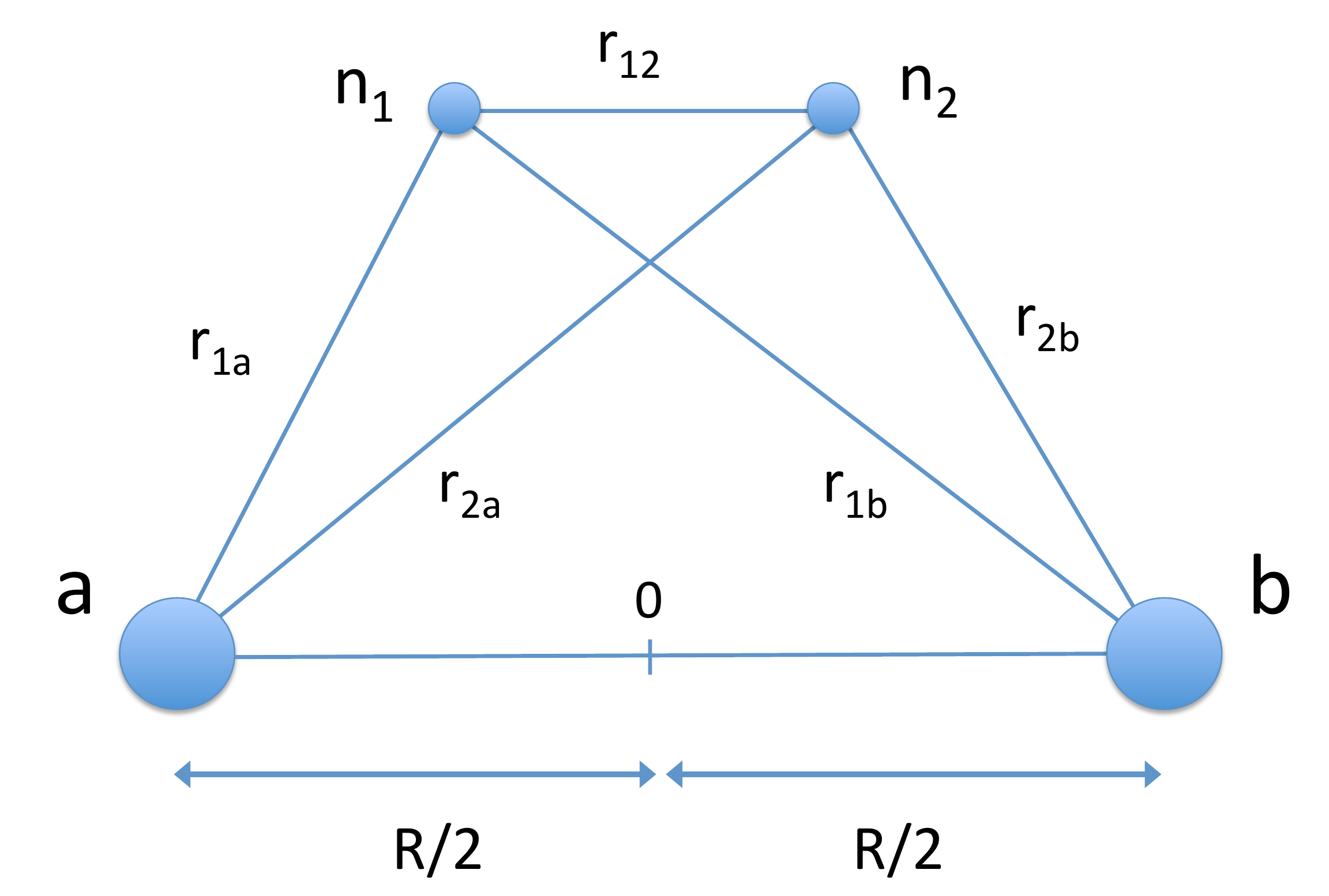}}\par}
\caption{\label{figm} Coordinates used in text.}
\end{figure}

Another clarification if a possible molecular state can exist in nuclear systems is obtained by using  the Bohr-Oppenheimer approximation for adiabatic systems. In this model, molecular states are built from the wave function tails of the two-body subsystems.  One starts with the total Hamiltonian for the three-body system  
 \begin{eqnarray}
{\cal  H}&=& -{\hbar^2\over 2 m}\nabla^2_1-{\hbar^2\over2 m}\nabla^2_2
+V_a (r_{1a})
+V_b(r_{1b)}\nonumber\\&+&V_a(r_{2a})+V_b(r_{2b})+v_{12}(r_{12}) \ , \label{hamil}
\end{eqnarray}
where $r_{ia}$ ($r_{ib}$) means the position of nucleon 
$i$ with respect to the core $a$ ($b$) and $r_{12}$ is the relative distance between the two neutrons (see Figure \ref{figm}).  Then, for the sake of simplicity and without loss of generality, we neglect the neutron-neutron interaction energy $\epsilon_{pair}$. A variational calculation for the energy of the system based on the assumption that the two-center wavefunctions are linear combinations of the two states centered in each of the two nuclei  separately, yields \cite{BO27}
\begin{equation}
E(R)={\epsilon_{2n}(1+{\cal O}^2)+2{\cal O} J+2I \over 1+{\cal O}^2},\label{BOA}
\end{equation}
where $\epsilon_{2n}$ is twice the energy of the  n+core system.  The matrix elements in Eq. \eqref{BOA}  are given by
\begin{eqnarray}
I(R)&=&\int |\psi({\bf r})|^2 V(|{\bf R}-{\bf r}|) d^3r , \nonumber \\
J(R)&=&\int \psi^*({\bf r}) V(|{\bf R}-{\bf r}|) \psi(|{\bf R}-{\bf r}|)d^3r ,\nonumber\\
{\cal O}(R)&=&\int \psi^*({\bf r}) \psi(|{\bf R}-{\bf r}|)d^3r ,\label{BO}
\end{eqnarray}
where $\psi$ is the relative wave function of the n+core system and $R$ is the distance between the c.m. of the two nuclei. 

For as Rydberg molecule, the integrals in Eq. \eqref{BO} are determined by using the single-particle wave functions for the neutrons in a large $n$ orbital,  or small separation energy, which for simplicity we assume to be a s-state.    We apply this method to the $^{10}$Be+n+n+$^{10}$Be system, consisting of  a molecular state in the $^{22}$O nucleus. Our choice for this system is to show that the existence of a Rydberg nuclear molecule is not farfetched and that they could exist under reasonable assumptions concerning the magnitude of nuclear forces and extreme halo states. 
 
\begin{figure}[ht]
{\par\centering \resizebox*{0.45\textwidth}{!} {\includegraphics{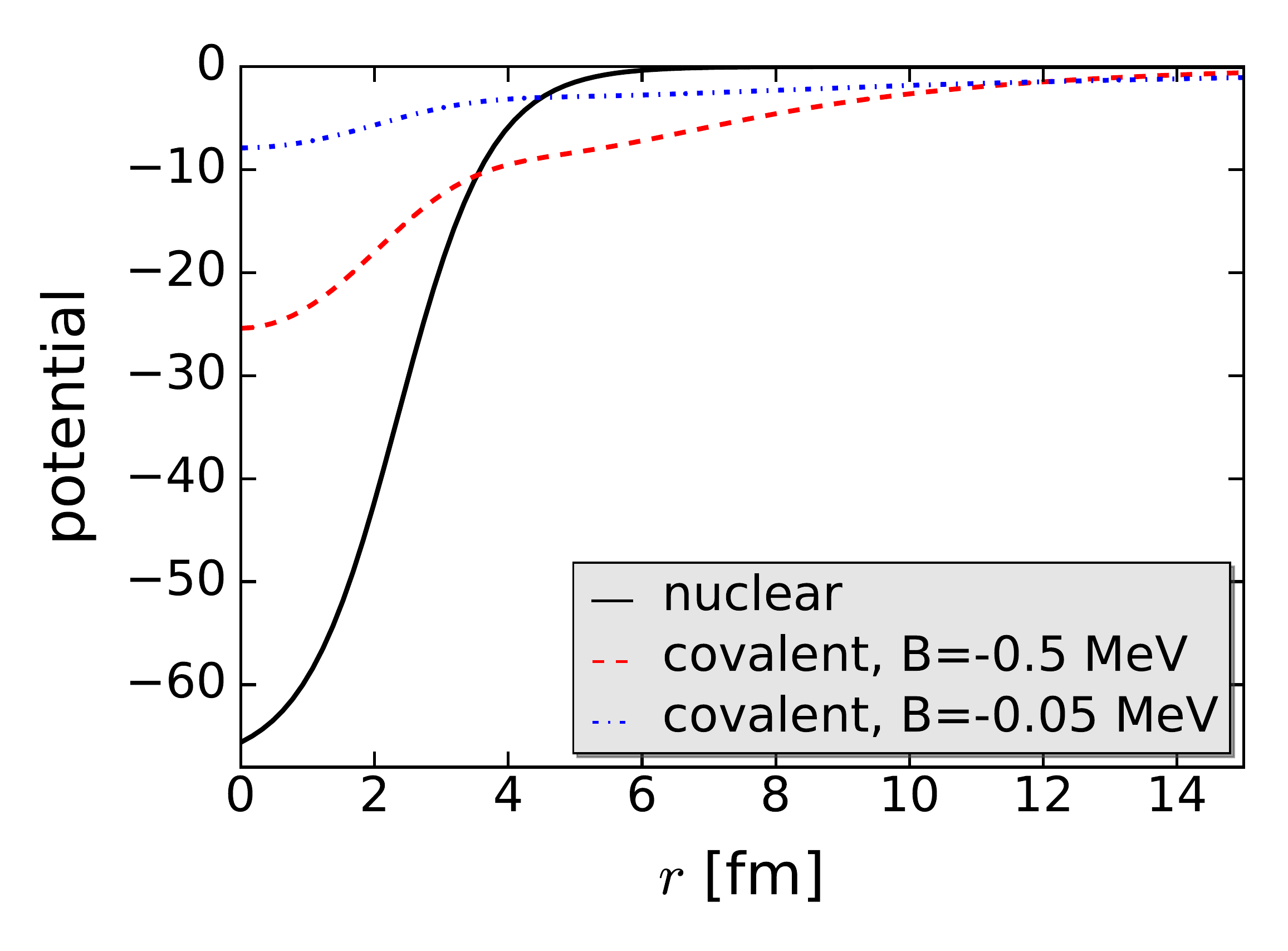}}\par}
\caption{\label{fig2} (Color online) The nuclear (solid line), and covalent potentials for the $^{10}$Be+n+n+$^{10}$Be system with neutron orbitals $2s_{1/2}$ and with the neutron binding energies of $B=-0.5$ MeV (dashed line) and $B=-0.05$ MeV (dotted line), respectively.}
\end{figure}

The energy given by the above equation minus the separation energy of the neutrons from $^{10}$Be+n+n+$^{10}$Be system, i.e. ${\cal E}_{covalent}(R)=E(R)-\epsilon_{2n}$, corresponds to the attractive energy, or molecular bond energy, caused by the exchange of the two neutrons between the $^{10}$Be cores.  For symmetric systems, $V_a=V_b$, the potential energy function for the whole system is obtained by adding the core-core nuclear and Coulomb potentials, 
i.e., 
$
V(R)=V_N(R)+{Z^2e^2/ R}+{\cal E}_{covalent}(R) 
$.
In Figure \ref{fig2} we plot the nuclear (solid line), and covalent potentials ${\cal E}_{covalent}$ for the $^{10}$Be+n+n+$^{10}$Be system with the neutron binding energies of $B=-0.5$ MeV (dashed line) and $B=-0.05$ MeV (dotted line), respectively. It is noticeable that the covalent bonding is non-negligible at large distances. The Coulomb repulsion between the nuclei is reduced by the molecular effect which can lower a resonance state to turn into a bound molecular state, specially for small binding energies. In fact, a similar molecular bonding model was proposed in Ref. \cite{Baha93} as a means to enhance fusion cross sections of halo nuclei at small energies as the molecular bonding effect decreases the potential height at the Coulomb barrier therefore increasing the barrier penetrability. Such effect has not yet been determined because of numerous other physical effects which can suppress or enhance   the fusion cross sections involving halo and neutron-rich nuclei \cite{Can15}. Contrary to the atomic systems, nuclei rarely display bound orbitals with large principal $n$ quantum number. For the  $^{10}$Be+n+n+$^{10}$Be covalent bonding we have used the $2s_{1/2}$ orbital which is far from being a large $n$ orbital. Nonetheless it displays the same physics leading to the formation of a covalent atomic bonding also increasing the possibility for the existence of a nuclear molecule. 
   
We now include the neutron-neutron interaction to calculate the ground-state of a nuclear molecule,  built from loosely-bound neutron+core states. This can be easily accomplished using a path integral variational  Monte-Carlo formalism (see, e.g., Refs. \cite{Cep95,Her82}). For the trial wave function in the variational method we assume again the validity of the Bohr-Oppenheimer approximation. Neglecting spin interactions, the neutrons are in antisymmetric spin-singlet states. Thus, the molecule wavefunction is symmetric in the neutron ${\bf r}_1$ and ${\bf r}_2$ coordinates, and the trial wavefunction does not need reference to spin variables, being of the form
\begin{eqnarray}
\Phi({\bf r}_1, {\bf r}_2, {\bf R})&=& \phi({\bf r}_1) \phi({\bf r}_2) f({\bf r}_{12}) \label{VMC}
\end{eqnarray}
where $\phi({\bf r}_1) = \phi({\bf r}_{1a})+\phi({\bf r}_{1b})$ with $ \phi({\bf r}_{1a})$ the single-particle wavefunction generated in the same way as we did before, but centered on the $a$ nucleus, while $ \phi({\bf r}_{1b})$ is centered on the $b$ nucleus (see Figure \ref{figm}). The correlation function $f$ is expected to be large when ${\bf r}_{12}$ is small and to decrease and become constant as the neutrons separate to large distances. Therefore, we assume a correlation function of the form $f({\bf r})=\exp\{-r/[r_0(1+\alpha r)]\}$ where we take $r_0=1$ fm and  $\alpha$ becomes a positive variational parameter, adjusted to get the minimum system energy, $E=\langle \Phi |{\cal H}|\Phi\rangle/\langle \Phi | \Phi \rangle$. This is achieved by sampling  $\Phi({\bf r}_1, {\bf r}_2, {\bf R})$ in a standard multidimensional Monte-Carlo integration procedure followed by a search for the minimum of $E$. 

The wave function for a given nuclear separation $R$, with the parameter $\alpha(R)$ obtained from the calculation described above is further  refined  in an additional variational procedure based upon the evolution of the imaginary-time Schr\"odinger equation by applying the operator $\exp(-{\cal H}t/\hbar)$ to the trial state $\Phi$ and considering the long-time limit $t \rightarrow \infty$. This  procedure is known as the Path Integral Monte Carlo (PIMC) method \cite{Cep95,Her82} and the wave function is written as
\begin{equation}
\Psi({\bf r}_1, {\bf r}_2, t)=  \exp\left[ \int_0^t E_c(t^\prime) d t^\prime/\hbar \right] e^{-{\cal H}t/\hbar} \Phi({\bf r}_1, {\bf r}_2), \label{pimcw}
\end{equation}
where $E_c(t)$ is identified with the ground-state energy 
\begin{equation}
\Sigma ={\langle \Psi |{\cal H}|\Phi\rangle\over \langle \Psi | \Phi \rangle}={\int d^3r_1d^3r_2 \Phi({\bf r}_1, {\bf r}_2)\Psi({\bf r}_1, {\bf r}_2, t)\epsilon({\bf r}_1, {\bf r}_2)\over \int d^3r_1d^3r_2 \Phi({\bf r}_1, {\bf r}_2)\Psi({\bf r}_1, {\bf r}_2, t)},
\end{equation}
when $t\rightarrow \infty$. The PIMC method is equivalent to solve numerically the path integral representation of Eq. \eqref{pimcw}. In the above expression $\epsilon({\bf r}_1, {\bf r}_2)=\Phi^{-1}({\bf r}_1, {\bf r}_2){\cal H}\Phi({\bf r}_1, {\bf r}_2)$ is the local energy. The wavefunction $\Phi$ approaches the (un-normalized) exact ground state  as t becomes large, as long as $\langle \Psi(t=0) |{\cal H}|\Phi\rangle \ne 0$. The total energy of the molecule is given by $E_{mol}(R)=\Sigma(R) + V_{Be-Be}(R) + Z_{Be}^2 e^2/R$. 

For the neutron-neutron interaction in Eq.  \eqref{hamil} we use  a simple radial square-well potential adjusted to give rise to an infinite scattering length for the n-n system. This is accomplished in practice by using a potential depth, $V_0$, in the limit of zero binding so that the depth and the potential  range $r_0$, are related by $V_0=-\pi^2 \hbar^2/2m_Nr_0^2$. Again, we make the arbitrary choice that $r_0=1$ fm. For the n+core potential we use the same potential model as described previously. Additionally, for  $^{10}$Be + $^{10}$Be  we choose a shallow Woods-Saxon potential yielding the lowest $s$ state at zero binding. We obtain for the depth $V_0=-9.04$ MeV, with radius $R=4.50$ fm and diffuseness $a=0.7$ fm. 

Our calculations show that at a nuclear ($^{10}$Be + $^{10}$Be) distance of 6 fm one finds molecular states with $E_{mol}=-0.698 \pm 0.04$ MeV, monotonically decreasing to zero and becoming unbound beyond $R\simeq 12$ fm. A list of values of $E_{mol}$ for several distances $R$ is given in Table \ref{tab1}. The errors are due the Monte-Carlo sampling average dispersions. These are (numerical) loosely bound systems which can easily disappear if details of the nuclear interactions are changed, even slightly. These results are for 2$s_{1/2}$ neutron orbitals with $E_B=-0.5$ MeV.  Using the 1$p_{1/2}$ state in $^{11}$Be at 320 keV excitation energy yields zero binding for any reasonable ($R\gtrsim 6$ fm) separation for the $^{10}$Be cores.  This shows that, at least for light nuclei, the existence of nuclear molecules depends on very fine tuning of the nuclear interactions and binding energies.  However, our result also point to possible molecular states which cannot be ruled out due to the covalent bonding effect inherent to nucleon exchange.

\begin{table}[t]
\caption{Total energy of the  $^{10}$Be + n + n + $^{10}$Be as a function of the distance between the $^{10}$Be cores.}
\begin{tabular}{cc|c}
\hline\hline
&  R [fm]      &  $E_{mol}$ [MeV]   \\
\tableline
&6 & $-0.689 \pm 0.04$   \\
&8 &  $-0.238 \pm 0.02$ \\
&10 & $-0.141 \pm 0.01$ \\
&12 & $-0.0095 \pm  0.01$ \\
& $\gtrsim 13$&unbound\\
\hline\hline
\label{tab1}
\end{tabular}
\end{table}
  
\begin{figure}[ht]
{\par\centering \resizebox*{0.45\textwidth}{!} {\includegraphics{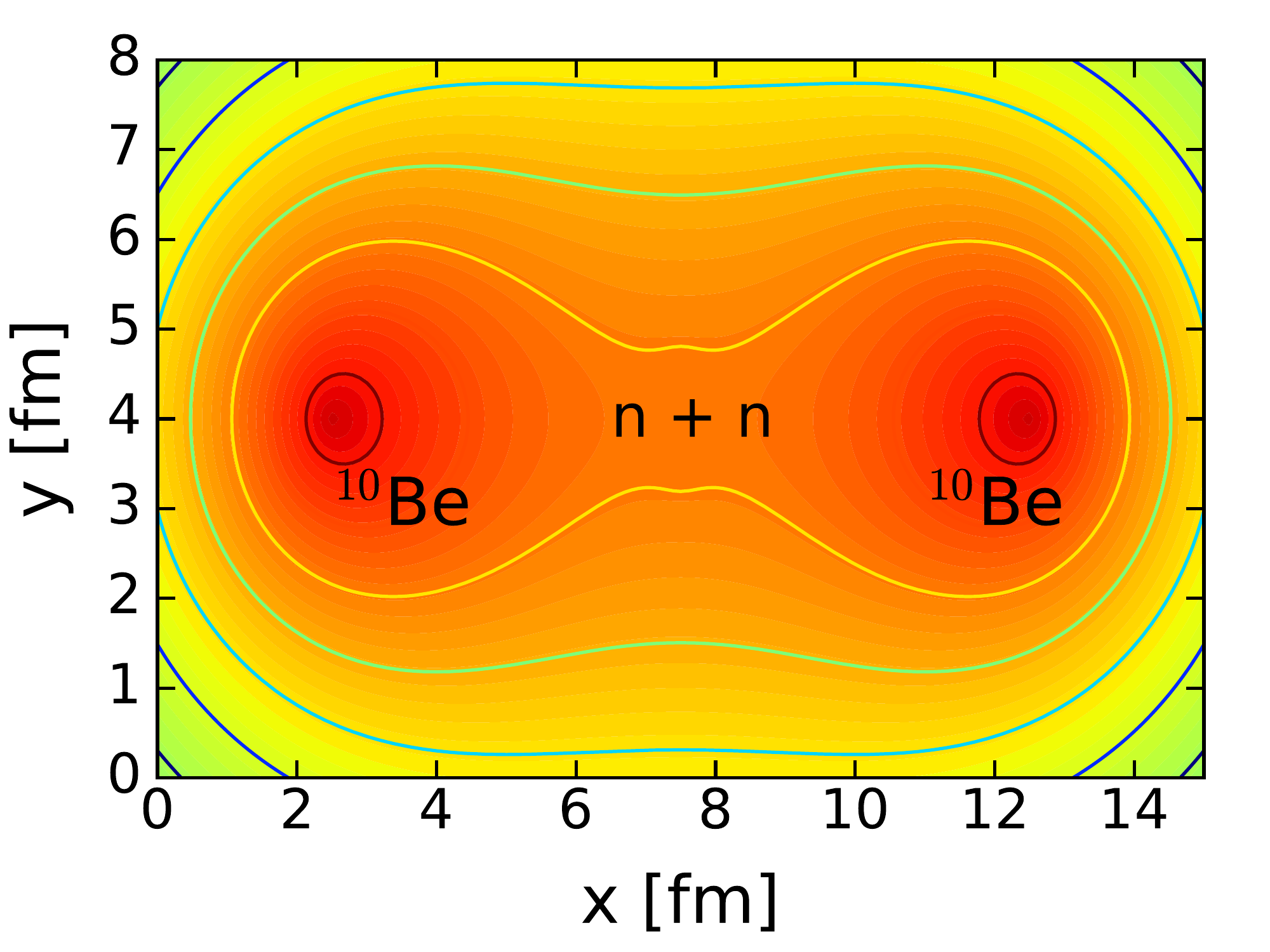}}\par}
\caption{\label{fig4} (Color online) Contour plot of the one-body matter density for a hypothetical $^{10}$Be+n+n+$^{10}$Be molecule with a separation distance of $R=10$ fm. The densities are projected onto a plane containing the major symmetry axis.}
\end{figure}

For the sake of clarity, we show in Fig. \ref{fig4} a contour plot of the one-body matter density for a hypothetical $^{10}$Be+n+n+$^{10}$Be molecule with a separation distance of $R=10$ fm. The densities are projected onto a plane containing the major symmetry axis. In this plot we use a variational Monte-Carlo wavefunction as in Eq. \eqref{VMC} with the variational parameter $\alpha=1.24$ obtained for the ground state. The results obtained with the PIMC are very similar. The two neutrons contribute to filling in the region between the two nuclei, as expected for the ground state of a covalent nuclear molecule. This situation seems to be completely different than that observed in other exotic nuclear systems as, e.g., the $^{11}$Li nucleus in which an opening angle between the lines joining the two neutrons and the $^9$Li core is observed \cite{BH07,HS07}. Here, the covalent bonding correlation prevails over other possible configurations. Such variety of correlations might  clarify many aspects of dilute neutron systems, such as the tetraneutron \cite{Mar02,BZ03,Kis16,BZ16} or a tri-neutron \cite{Gan17} system which have attracted a large interest by the nuclear physics community because of their largely unpredicted existence decades ago \cite{BZ03,BZ16,Fos17,Gan17,Laz17}.

A comment about the stability of this fragile system.
The neutron separation energy in the $^{11}$Be is 0.5 MeV, therefore if the
binding energy of $^{10}$Be+n+n+$^{10}$Be molecule would be less than 0.5 MeV,
the molecule can decay into $^{11}$Be and n+$^{10}$Be or $^{11}$Be+$^{11}$Be. What can turn less probable the decay is the size of the molecule and its two-neutron halo and in addition the large healing distance of the  halo neutron in $^{11}$Be: R$\sim\sqrt{41.47 \ {\rm MeV \ fm}^2/0.5 \ {\rm MeV}}\approx$ 9 fm, obtained from the neutron separation energy in $^{11}$Be.
The effective $^{10}$Be-$^{10}$Be potential has a shallow and long tail,  being attractive with strength   running from $-0.6$ MeV  to 0 for  $6\ {\rm fm}  < R < 13$ fm  (see Table I).
It could eventually accommodate a weakly bound large
molecule, which could reach a size with an internuclear distance extending beyond 12 fm. More specifically, we can consider for
example the decay amplitude, A, of the molecule into $^{11}$Be+ $^{11}$Be which is schematically
given by 
\begin{equation}
A = \left<\psi_{(^{11}Be)}\psi_{(^{11}Be)} \left| V_{nn}\right| \Psi_{(Molecule)}\right>. 
\end{equation}
As $V_{nn}$ is of short
range, $\sim1$ fm, and the average distance between the neutrons in the molecule
could be even larger than $6 -12$ fm, together with the halo neutron in each $^{11}$Be
being distant from the core ($\sim 9$ fm) it would lead to an overlap of very dilute systems of neutrons, and therefore damps the decay amplitude and suppresses the coupling to the decay channel. A similar argument could be used for the three-body decay channel Molecule $\rightarrow \ ^{11}$Be + n + $^{10}$Be, which could also be suppressed. This argument suggests that the stability of the molecule against decay can become reasonably robust as the binding energy decreases.

We also venture to suggest that fragile systems such as the Rydberg nuclear molecule could dissociate as a result of the influence of an external parameter, such as the temperature in the interior of stars. On the other hand, in the crust of neutron stars the opposite happens, as we have learned from the formation of nuclear molecules in the pasta configuration.

In conclusion, we have studied the possible existence  of a ``Rydberg" nuclear molecule composed of two $0^+$ nuclei and two neutrons shared covalently between them. We have used three methods that emphasize two neutrons with similar configurations as electrons in a $H_2$ molecule. A covalent bond appears in the system which may lead to a bound molecular state. Our calculations are simplified but also consistent with our present knowledge of basic facts of the nucleon-nucleon interaction. We have neglected several effects which might become important in the fine tuning aspect of the molecular binding, such as spin and nuclear recoil.  Our arguments are based on the Bohr-Oppenheimer approximation which assumes that the neutron orbitals adjust rapidly to the much slower motion of the nuclear cores.  We have used the $^{10}$Be + n + n + $^{10}$Be molecular system as a playground, because some of the known states  of $^{11}$Be can be used as a starting point to test the effect of the wavefunction tails. 

Due to the stronger Coulomb repulsion of the nuclei it is unlikely that heavier systems would become bound by such a covalent mechanism, except perhaps at much larger nuclear separations, which could lead to a proper Rydberg nuclear molecule. Our calculations show that one cannot confirm, but it is also difficult to rule out the existence of such molecules based on what is known about nuclear interactions. They might exist in rich neutron environments within stellar sites, or maybe even created in laboratory as byproduct of nuclear fragmentation. Another possibility is to gather a large number of, say, $^{11}$Be, in a trap and monitor the number of these halo (Rydberg) nuclei, and the very elongated Rydberg molecule $^{22}$O (ground state half life 2.25 s).  The experimental proof of the formation of a Rydberg nuclear molecule involving a dimer of two halo nuclei, extends the study of clustering to exotica. An experimental verification of their existence would allow for a wonderful laboratory for the complexity of nuclear wave functions at large distances, with observables arising from their implications for feeble but probably rich vibrational and rotational structures. 

{\it Acknowledgements.} 
This work s supported by the U.S. DOE grants DE-FG02-08ER41533 and the U.S. NSF Grant No. 1415656 (CAB), and by the Brazilian agencies, Funda\c c\~ao de Amparo \`a Pesquisa do Estado de S\~ao Paulo (FAPESP), the  Conselho Nacional de Desenvolvimento Cient\'ifico e Tecnol\'ogico  (CNPq). CAB also acknowledges a Visiting Professor support from FAPESP and from HIC for GAIR, and MSH acknowledges a Senior Visiting Professorship granted by the Coordena\c c\~ao de Aperfei\c coamento de Pessoal de N\'ivel Superior (CAPES), through the CAPES/ITA-PVS program.


\begin{thebibliography}{4}
\bibitem{Ga94} T. Gallagher, ``Rydberg Atoms", Cambridge University Press (1994). 
\bibitem{Ja00} D. Jaksch et al.,  ``Fast Quantum Gates for Neutral Atoms", Phys. Rev. Lett. {85},   2208 (2000).
\bibitem{Gr00} C.H. Greene, A.S. Dickinson and H.R. Sadeghpour,  ``Creation of Polar and Nonpolar Ultra-Long-Range Rydberg Molecules", Phys. Rev. Lett. {85}, 2458 (2000).
\bibitem{Ov09} K. R. Overstreet, A. Schwettmann, J. Tallant, D. Booth, J. P. Shaffer, ``Observation of electric-field-induced Cs Rydberg atom macrodimers" Nature  Physics {5}, 581 (2009).
\bibitem{Ve09} Vera Bendkowsky, Bj\"orn Butscher, Johannes Nipper, James P. Shaffer, Robert L\"ow, Tilman Pfau, ``Observation of ultralong-range Rydberg molecules", Nature {458}, 1005 (2009).
\bibitem{BRM1} Thomas NiederprŸm, Oliver Thomas, Tanita Eichert, Carsten Lippe, Jesœs PŽrez-R'os, Chris H. Greene, Herwig Ott, ``Observation of pendular butterfly Rydberg molecules", Nature Communications 7, 12820 (2016).
\bibitem{BCH93} C.A. Bertulani, L.F. Canto and M.S. Hussein, ``The structure and reactions of neutron-rich nuclei,", Phys. Rep. {226}, 281 (1993). 
\bibitem{Ef70} V. Efimov, ``Energy levels arising from resonant two-body forces in a three-body system", Phys. Lett. B {33}, 563 (1970).
\bibitem{Fre12} T. Frederico, A. Delfino, L. Tomio, M.T. Yamashita,  ``Universal aspects of light halo nuclei ", Prog. Part. Nucl. Phys. 67, 
939 (2012).
\bibitem{Pla04} L. Platter, H.-W. Hammer, Ulf-G. Meissner, ``Four-boson system with short-range interactions", Phys. Rev. A70, 052101 (2004). 
\bibitem{Yam06}M. T. Yamashita, L Tomio, A Delfino, T Frederico, ``Four-boson scale near a Feshbach resonance", Europhys. Lett.  75, 555 (2006). 
\bibitem{St09} J. von Stecher, J. P. D'Incao  and Chris H. Greene, ``Signatures of universal four-body phenomena and their relation to the Efimov effect", Nature Physics {5}, 417 (2009).
\bibitem{Had11} M. R. Hadizadeh, M. T. Yamashita, L. Tomio, A. Delfino, T. Frederico, ``Scaling Properties of Universal Tetramers", Phys. Rev. Lett. 107, 135304 (2011).
\bibitem{Fer09} F. Ferlaino, S. Knoop, M. Berninger, W. Harm, J. P. D'Incao, H.-C. NŠgerl, R. Grimm, ``Evidence for Universal Four-Body States Tied to an Efimov Trimer", Phys. Rev. Lett. 102, 140401 (2009). 
\bibitem{Fe36} E. Fermi,  ``Motion of neutrons in hydrogenous substances", Ricerca Scientifica {7}, 13 (1936).
\bibitem{BO27} M. Born, J.R.  Oppenheimer, ``Zur Quantentheorie der Molekeln", Annalen der Physik {389}, 457 (1927).
\bibitem{Baha93} C.A. Bertulani and A.B. Balantekin, ``Molecular bonding effects in the fusion of halo nuclei", Phys. Lett. B {314}, 275 (1993).
\bibitem{Can15} L.F. Canto, P.R.S. Gomes, R. Donangelo, J. Lubian, M.S. Hussein, ``Recent developments in fusion and direct reactions with weakly bound nuclei", Phys. Reports 596, 1 (2015). 
\bibitem{Cep95} D.M. Ceperley, ``Path-integral computation of the low-temperature properties of liquid He-4", Rev. Mod. Phys. 67, 279-355 (1995).
\bibitem{Her82} M.F. Herman, E.J. Bruskin, and B.J. Berne, ``On path integral Monte Carlo simulations", J. Chem. Phys. 76, 5150 (1982).
\bibitem{BH07} C.A. Bertulani and M.S. Hussein, ``Geometry of Borromean Halo Nuclei", Phys. Rev. C76, 051602(R) (2007).
\bibitem{HS07} K. Hagino and H. Sagawa,  ``Dipole excitation and geometry of borromean nuclei", Phys. Rev. C76 047302 (2007).
\bibitem{Mar02} F.M. Marqu\'es, et al., ``The Detection of neutron clusters", Phys. Rev. C 65, 044006 (2002).
\bibitem{BZ03} C.A. Bertulani and V.G. Zelevinsky, ``Tetraneutron as a dineutron-dineutron molecule", J. Phys. G 29 (2003) 2431.
\bibitem{Kis16} K. Kisamori, et al., ``Candidate Resonant Tetraneutron State Populated by the  4He(8He,8Be) Reaction", Phys. Rev. Lett. 116, 052501 (2016).
 \bibitem{BZ16} C.A. Bertulani and V. Zelevinsky, ``Four neutrons together momentarily", Nature 17884 (2016).
 \bibitem{Fos17} K. Fossez, J. Rotureau, N. Michel, and M. Ploszajczak, ``Can tetraneutron be a narrow resonance?", to be published (2017).
\bibitem{Gan17} S. Gandolfi, H.-W. Hammer, P. Klos, J. E. Lynn, and A. Schwenk, ``Is a trineutron resonance lower in energy than a tetraneutron resonance?", to be published (2017).
\bibitem{Laz17} R. Lazauskas, J. Carbonell and E. Hiyama, ``Modeling double charge exchange response function for tetraneutron system", to be published (2017).
\end{thebibliography}
\end{document}